\documentclass[twocolumn,showpacs,amssymb,prd]{revtex4}

\usepackage{graphicx}
\usepackage{dcolumn}
\usepackage{bm}
\usepackage{epsfig}

\begin{document}
\newcommand{\gsim}{\hbox{\rlap{$^>$}$_\sim$}}
vv\newcommand{\lsim}{\hbox{\rlap{$^<$}$_\sim$}}

\title{Are Fast Radio Bursts Produced By Large Glitches\\
                Of Anomalous X-ray Pulsars?} 

\author{Shlomo Dado,  Arnon Dar}\affiliation{Physics Department, Technion, 
Haifa, Israel}

\begin{abstract} 
Starquakes and internal phase transitions within anomalous x-ray 
pulsars (AXPs) and soft $\gamma$-ray repeaters (SGRs) can produce mini 
contractions and pulsar glitches. Shocks break out from their 
surface following such contractions produce thermal x/$\gamma$-ray 
bursts. Highly relativistic dipolar $e^+e^-$ bunches launched from the 
pulsar polar caps emit fast radio bursts (FRBs) of narrowly beamed 
coherent curvature radiation, visible if they point in the direction 
of Earth.  Although these surface x/$\gamma$-ray bursts are 
isotropic and are many orders of magnitude more energetic than the FRBs, 
they are detectable by the current all sky x-ray and $\gamma$-ray 
monitors only from our galaxy and very nearby galaxies.
\end{abstract}


\maketitle

\section{Introductin}
Fast radio bursts (FRBs) are short radio pulses of length ranging 
from a fraction of a millisecond (ms) to a few ms from extragalactic 
sources [1]. They were first discovered in 2007 [2]. Their 
extragalactic origin was indicated by their dispersion measures 
and by their isotropic distribution in the sky [1]. 
Their estimated distances from their dispersion measure and  
radio fluence implied an energy release by an isotropic FRB in GHz 
radio waves roughly between $10^{37}$ erg and $10^{42}$ erg [1].

By the beginning of 2016, the lack of repeating pulses among FRBs 
known at that time, despite hundreds of hours of follow-up time 
[2,3], led to the wide spread belief that FRBs are one-time 
events. However, FRB121102 [4], was followed by many more FRBs 
from the same source. Additional constraints on the nature of 
FRB 121102 were provided in 2017 by the precise localization of 
its source using the Karl Jansky Very Large Array [5]. Radio 
observations using the European Very Long Baseline Interferometry 
Network and Arecibo provided compelling evidence for its 
positional association with a low-metalicity star-forming dwarf 
galaxy at a redshift z=0.192 [5]. This redshift, corresponding 
to a luminosity distance of $\approx$ Gpc, was consistent with 
that obtained before from its dispersion measure, which supported 
the estimated isotropic equivalent energy release of cosmological FRBs 
being between $10^{37}$ erg and $10^{42}$ erg [1] in the radio band.

Moreover, FRB 180814 [6] and 8 additional FRB sources discovered 
last year by the Canadian Hydrogen Intensity Mapping Experiment 
(CHIME) [7] were found to be repeaters, and 16.3 days 
periodicity was found in the repeating activity of  
FRB 180916 observed with the CHIME telescope [8]. They ruled 
out the possibility that cataclysmic events, such as stellar 
explosions or stellar mergers, are a common source of all types 
of FRBs [9]. However, they have not ruled out the possibility that 
single event FRBs and repeaters do have a common origin.

The sub-ms rise time of FRB pulses plus causality implies  that 
RBBs are produced by very compact sources such as pulsars. Further 
indication of a pulsar origin of extragalactic FRBs is an average 
pulse shape (after correcting for dispersion) similar 
to that of radio pulsars [2]. But, perhaps the strongest evidence so 
far for a possible FRB-pulsar association came on April 28, 2020. A 
double peak FRB 200428 was detected [10] from the direction of the 
Galactic soft gamma ray repeater 1935+2154, which coincided in time 
(after correcting for dispersion) with a double spike x-ray flare 
from that source [11]. Although FRB 200428 was a thousand times less 
bright than typical extragalactic FRBs [10], it raised the 
possibility that both Galactic and extragalactic FRBs are produced by 
soft gamma ray repeaters (SGRs) and anomalous x-ray pulsars (AXPs) 
[10]. Probably, only extragalactic FRBs that point very near the direction 
of Earth, are detectable from large cosmological distances, while 
Galactic FRBs, because of their proximity, are detectable up to much 
larger viewing angles. As in the case of gamma ray bursts (GRBs),
the strong dependence of the equivalent 
isotropic energy and luminosity of FRBs on their viewing angle 
yields, on average, much larger isotropic equivalent energy, peak 
energy, and peak luminosity, of extragalactic FRBs compared to those 
of Galactic and very nearby FRBs [12].
 
Moreover, large pulsar glitches have been seen to coincide within errors 
with giant x/soft gamma-ray flares of Galactic SGRs/AXPs [13]. Such 
glitches may have been produced by mini contractions of pulsars 
following starquakes or internal phase transitions [14]. They raise the 
possibility that such mini contractions lead to shocks break out from 
the surface of pulsars. Such shocks break-out in SGRs/AXPs, although much 
weaker, are analogous to those observed in core collapse (CC) supernova 
explosions (SNe) of massive stars [15]. However, in SGRs/AXPs, such 
shocks break-out are expected to occur very shortly after the mini 
contraction. This is because of the much smaller size and much shorter 
dynamical time scales within pulsars due to their enormous density 
compared to those of massive stars. Shocks from starquakes in the crust 
layer of pulsars can reach the surface directly and produce a hot area 
on the surface of the pulsars. Mini contraction following internal phase 
transition can produce much stronger shocks which can be reflected from 
the center of the pulsar and break out from its entire surface.

By now, improved estimates of the distance to several SGRs, allow 
critical tests of whether the x/$\gamma$-ray emission from 
large flares of SGRs are consistent with being thermal radiation from 
the surface of neutron stars. Indeed, as will be shown below, the 
record giant flare observed so far from an SGR, i.e., that of SGR 
1806-20 on 27 December 2004 [16], and the intermediate x/soft 
$\gamma$-ray flare observed on 12 April 2005 from SGR 1935+2154 [17] 
are consistent with being thermal emissions from the entire surface 
of a canonical neutron star. In all other cases where such tests 
were possible, the emitting surface area was 
equal or smaller than that of a canonical neutron star.

All together, the above seems to suggest that FRBs are produced by large 
glitches of SGRs/AXPs in external galaxies. But, as we shall show below, 
only a small fraction of the total gravitational energy release in SGR 
glitches produced by mini contractions, is used to spin up these 
pulsars. Although most of the released energy is emitted as a burst of 
isotropic thermal x/soft $\gamma$-rays, it is detectable by the 
current all sky x/$\gamma$-ray monitors only from SGRs within our galaxy 
and nearby galaxies. However, part of the released energy in glicthes is 
emitted as a coherent curvature radiation from highly relativistic 
dipolar $e^+e^-$ plasmoids launched from the magnetic poles along the 
direction of the pulsar magnetic moment [18]. Such curvature radiation 
in the radio band is beamed mainly along the initial direction of motion 
of the dipolar jets. If such a jet happens to point in/near the 
direction of Earth, it becomes detectable by the largest radio 
telescopes/arrays up to very large cosmological distances as an FRB with 
characteristic pulse shape, peak energy and a very large linear and much 
smaller circular polarization [19].

\section{FRBs From  Pulsar Glitches.}
A pulsar glitch [20] is a sudden increase in the pulsar's rotational 
frequency, which usually decreases steadily due to braking provided 
by the emission of radiation, winds and high-energy particles. The 
exact cause of such glitches is still unknown. The prevailing view 
is that they are caused by an internal process within the pulsars 
such as an increase in the pulsar's crust rotational frequency by a 
brief coupling of an hypothesized pulsar's faster-spinning superfluid 
core [21] to the crust, which are usually decoupled. This brief 
coupling transfers angular momentum from the core to the crust
of the pulsar which causes an increase in its observed  
rotational frequency [22].

An alternative hypothesis for the origin of pulsar glitches is near 
surface starquakes/internal phase transitions which involve a sudden 
gravitational contraction of the pulsar that decreases its moment of 
inertia and speeds up its rotation within a very short time. The 
relatively small size of neutron stars and their very high density $\rho$ 
yield a dynamical time scale $\sim\!1\!/ \sqrt{G\rho}\!\sim\! 0.1$ ms, 
for mini contractions, which can explain the observed short pulse 
duration of FRBs. A large angular momentum may suppress contraction in 
fast rotating pulsars. It may explain why large glitches are much more 
prevailing in the slowly rotating SGRs/AXPs than in ordinary pulsars, and 
are extremely rare in ms pulsars [23].\\

\noindent
{\bf Shock break out flares in SGRs/AXPs?} A sudden mini contraction of a 
slowly rotating pulsar, following a starquake or an internal phase 
transition, may produce a shock wave, which converges towards the center 
and reflected back towards the surface. Like in core collapse supernovae 
explosions, the shock break out from the surface of the SGR/AXP is expected 
to produce a flash of radiation [15]. The finger prints of such a shock 
break out flash from the surface of a pulsar are a black body spectrum and 
a surface area consistent with that of a neutron star. Although the 
spectral energy density was reported for several giant flares of SGRs, only 
in two of these cases the distance to the SGRs/AXP by now are known well 
enough to allow a critical test of whether the lightcurve of the flare 
shows evidence for a shock break out from a pulsar. They include the giant 
flare of SGR 1806-20 on 27 December 2004 [16] and the large burst of SGR 
1935+2154 on 12 April 2005 [17].\\

\noindent
{\bf The giant flare of SGR 1806-20 on 27/12/2004} 
had an initial spike of a width $W\!\approx\!0.125$ s, a total energy
$E(spike)\!\approx \!(1.2\pm 0.3)\times 10^{46}$ erg [16]
assuming isotropic emission at a distance 
$D\!=\!8.7\!+\!1.8/\!-1.5$ kpc [24], and a black body like 
spectrum with a peak temperature $T\!\approx 265\!\pm\! 15$ keV [16]. 
Thus, the radius of the emitted source was 
\begin{equation}
R\!\approx \! \left[{E{spike}\over 4\pi\sigma T^4 W}\right]^{1/2}
   \!\approx\!12.3\!\pm\! 2.3\,  {\rm km},  
\end{equation}
consistent with that of a canonical neutron star.\\

\noindent 
{\bf The intermediate flare of SGR 1935+2154 on 12/4/2005}  
had a double peak structure [17].
The first peak had a black body spectrum with a temperature 
$T\!=\!6.4\!\pm\!0.4$ keV. The 
assumption of isotropic emission at a distance $10$ kpc  
has yielded [17] $R\!\approx\!\sqrt{455\!+\!73/\!-\!55}\!\approx \! 
21\!\pm\!2$ km. However, recently the  distance to 
SGR 1935+2154 has been estimated to be only [25] $6.6\!\pm\! 0.7$ kpc,
yielding $R\!\approx\!14\!\pm\!3$ km, consistent with that of a 
canonical neutron star. \\

\noindent 
{\bf Energy release in pulsar glitches.}
Consider an SGR/AXP with a 
canonical neutron star properties; a mass $M\!\approx\!1.4\,M_\odot$, a 
radius $R\!\approx\! 10$ km, a period $P$ (rotational frequency 
$\nu\!=\!1/P$), and a moment of inertia $I\!\approx\!(2/5)\,M\,R^2\!\approx 
\! 1.12\times 10^{45}\,{\rm gm\,cm^2}$, whose radius contracts in a major 
glitch by $\Delta R$. Angular momentum conservation, 
\begin{equation} 
\Delta L\!\approx 2\pi\,I \Delta \nu\!+ 2\pi\,\Delta I\,\nu\!=\!0, 
\end{equation} yields 
\begin{equation} 
2\,\Delta R/R = -\Delta\nu/\nu\, 
\end{equation} 
and a rotational energy increase, 
\begin{equation}
\Delta E_{rot}\!=\!(\Delta\nu/\nu)E_{rot} 
\end{equation} 
in such a glitch. The sudden contraction of the pulsar is 
accompanied by a gravitational energy release, 
\begin{equation} 
\Delta E_g\!\approx\!(3\,G\,M^2/5R)(\Delta R/R) 
           \!\approx\!(E_g/2)(\Delta\nu/\nu)\,, 
\end{equation} 
where $G$ is the gravitational constant. In SGRs/AXPs with a typical 
period $P\gsim 1$ s, the gravitational energy release in a glitch is by far 
larger than the increase in their rotational energy,
\begin{equation} 
{\Delta E_g \over \Delta E_{rot}}\!\approx \!{3G\,M\,P^2\over 4\pi^2 R^3}
      \!\approx\!1.4\times 10^7\!(P/{\rm s})^2. 
\end{equation}
According to the virial theorem, half the gravitational energy release is 
converted to internal kinetic energy, part of which is used to increase the 
millisecond pulsar (MSP) rotational energy. However, since the gravitational 
energy in SGRs is 
much larger than the rotational energy, eq.(6) implies that $\Delta E_{rad}$, 
the radiated energy from major glitches in SGRs/AXPs, is bounded roughly by 
\begin{equation} 
\Delta E_{rad}\!\leq\! 
(3/20)(GM^2/R)\Delta \nu/\nu\,. 
\end{equation} 
The largest glitches observed so far in SGRs/AXPs had 
$\Delta \nu/\nu\!\leq\!10^{-5}$. For such glitches, eq.(7) yields $\Delta 
E_{rad}\approx 8\times 10^{47}$ erg. Probably, the bulk of this energy,
escapes as a short burst of neutrinos, like in core collapse SNe, 
followed by a short flash of surface thermal  x/$\gamma$-rays.
However, because of the very             
small radius and the huge mean density of pulsars relative to 
those of massive stars, the short spike of thermal x/$\gamma$-ray 
surface radiation from a shock break out following a pulsar glitch,
can even precede the neutrino burst. \\

\section{FRB- coherent curvature radiation?}
The main observed properties of FRBs are those expected 
of narrowly beamed coherent curvature radiation [26]
emitted by SGRs/AXPs following large glitches.\\

\noindent
{\bf Spectrum}. A characteristic frequency of the curvature radiation 
emitted by a bunch of highly relativistic electrons moving with a bulk motion  
Lorentz factor $\Gamma\!\gg\!1$ along a track with a curvature radius 
$\rho_c$ was defined as [26], 
\begin{equation}
\nu_c\!=\!3c\Gamma^3/4\pi \rho_c\,.
\end{equation}
The spectral distribution of the radiated energy, $dW/d\nu$, has the 
standard synchrotron radiation spectral distribution [26] which, in
vacuum, is a function of the ratio $x=\nu/\nu_c$ . 
In the pulsar rest frame, to a good approximation $dW/d\nu\!\propto\! 
x^{1/3}$ well below its peak value at $x=0.29$, and changes to  
$dW/d\nu\!\propto\sqrt{x}e^{-x}$ well above it.\\

 \noindent
{\bf Beaming}. The curvature radiation  from  highly relativistic electrons
moving along a curved magnetic field line is collimated into a narrow 
cone of opening angle $\approx \!1/\Gamma$ along their direction of motion. 
Eq.(8) and the locally observed FRB peak frequencies  around 1.5 MHz, which  
satisfy $\nu_p\!\approx\!0.29\,\nu_c/(1+z)\!\sim\!1.5(1\!+\!z)$ GHz, 
imply $\left <\Gamma\right >\!\approx\! 90(1+z)^{1/3}$. \\

\noindent
{\bf Pulse shape of FRB}. If FRBs and ordinary pulsar pulses are 
produced by curvature radiation, then, 
{\it after correcting for dispersion 
and redshift}, FRBs are expected to have a pulse shape similar to 
the average pulse shape of Galactic radio pulsars. The fast expansion 
of the plasmoids and the decline of the energy density of the magnetic 
field with increasing distance from the pulsar yield a
FRED (fast rise, exponential decay) energy fluence with a shape similar 
to that of GRBs pulses [12]
\begin{equation}
F(t)\propto [t^2/(t^2\!+\!\Delta^2)]^{2\alpha} e^{-\beta t},
\end{equation}
where $\alpha$, $\beta$  and $\Delta$, are constants, which
vary between different FRBs, and $t$ is the time since the 
beginning of the pulse.\\

\noindent
{\bf Polarization}. The curvature radiation is strongly polarized in the plane 
cf curvature. As the radio beam sweeps across the line of sight, 
the plane of polarization rotates up to 180 degrees [26]. \\

{\bf Periodic FRB activity ?} Pulsars in highly eccentric orbits 
around a massive star in compact binaries may suffer periodic glitches 
triggered by mass accretion episodes which take place mainly 
near perihelion. Such activity can yield semi-periodic  
FRB activity with a period  equal to the orbital period 
of the MSP around the massive star.
\begin{equation}
T=\left[{4\pi^2\over \sqrt{G(M_{n*}\!+\!M_*)}}\right]^{1/2}
  \left[R_p\over(1-\epsilon)\right]^{3/2} 
\end{equation} 
where $M_{n*}$ and $M_*$ are, respectively, the masses of the 
pulsar and the 
massive star in the compact binary, and $\epsilon$ and $R_p$ are, 
respectively, the eccentricity and the perihelion distance of the 
MSP orbit around the massive star. In the case of the periodic FRB 
180916, the observed period of its FRB activity was 
$T=16.3$ days [8].\\

\noindent 
{\bf The redshift distribution of FRBs.} 
The birth rate of SGRs/AXPs is a constant fraction of the birth rate 
of neutron stars in core collapse supernova explosions of short lived 
massive stars, which traces the star formation rate. The very small 
characteristic age $\tau\!=\!P/2\dot{P}$ of SGRs/AXPs compared to 
that of ordinary pulsars and the similar 
beaming of FRBs and long duration gamma ray bursts (LGRBs) imply that 
their production rates as a function of redshift are roughly 
proportional.\\

\noindent
{\bf Threat to life.} 
FRBs do not threat life on nearby life supporting planets. But 
large glitches of AXPs/SGRS which produce giant x/$\gamma$-ray 
flares do, independent of whether or not they were accompanied by 
an FRB. Although the electromagnetic and kinetic energy release in 
giant flares of AXPs/SGRs is much smaller than that released in SN 
explosions, the much higher rate of their giant flares compared to 
the birth rate of AXPs/SGRs, their very short duration, and their 
much harder radiation, produce a more serious threat to life on 
nearby life supporting planets. as was noticed in [27].

\section{Discussion and Conclusions.} 
The possibility that giant x/$\gamma$-ray flares of AXPs/SGRs, which are 
widely believed to be slowly rotating magnetars -highly magnetized, 
slowly rotating pulsars with a surface magnetic field in excess of 
$10^{14}$ Gauss [28]- are the main source of extragalactic FRBs has 
been raised recently by the discovery of FRB coincident with an x-ray 
flare from the Galactic SGR 1935+2154 [11]. The power supply for both 
a steady x-ray emission and x-ray flares, which exceeds by far the 
observed rotational energy loss of AXPs/SGRs, was claimed to be 
provided by the decay of their magnetic field energy [28]. If the 
rotational energy loss of such pulsars is entirely by magnetic dipole 
radiation, then their magnetic field at the equator satisfies 
$B\!\geq\!3\times 10^{19}\sqrt{P\dot{P}/s}$ [29], which has been 
widely adopted in estimating their dipole magnetic field [28].  
Indeed, a magnetic field energy of the order of $B^2R^3/10$, [29] 
could power both their steady isotropic x-ray emission and their 
giant flares, while their FRBs could be produced by short emission of 
highly beamed coherent curvature radiation. However, as discussed in 
the Appendix, despite the wide belief that AXPs are magnetars, which 
spin down by magnetic dipole radiation, and the decay of their 
magnetic field energy powers their x/$\gamma$-ray radiation and 
flares [28], there is no solid evidence in support of these 
assumptions (see the Appendix).

In this paper we have suggested a mechanism by which SGRs/AXPs produce narrowly 
beamed FRBs. We proposed that:\\ 
(a) the main power supply of AXPs is by a gravitational energy release in a slow 
contraction, rather than by the decay of an hypothetical ultra strong magnetic 
field [28],\\ 
(b) their spin down is dominated by emission of high energy charged cosmic ray 
and wind particles escaping along open magnetic field lines, rather than by magnetic 
dipole radiation,\\ 
(c) the gravitational energy release by a steady slow contraction powers their 
steady x-ray emission,\\ 
(d) sudden short increase in this rate following a crustal starquake or an internal phase 
transition produces a glitch and shock waves whose surface break-out powers a short 
thermal X/$\gamma$-ray flare and emission of bunches of $e^+e^-$ from their polar caps,\\
(e) such bunches produce FRBs - narrowly beamed short bursts of coherent curvature 
radiation in the MHz radio band, which are detectable only when they point in the 
direction of Earth,\\
(f) beamed FRBs can explain why no FRB has been detected before from the giant, 
presumably isotropic, X/$\gamma$-ray flare of SGR 1806-20 on 27 December 2004 [30], or in 
large X/$\gamma$-ray flares of other Galactic and nearby AXPs/SGRs, if they happened to 
be in the field of view of a large radio telescope.\\

So far no x/$\gamma$-ray flares and/or FRBs associated with glitches of Galactic pulsars 
other than SGRs/AXPs were reported. Strong centrifugal barriers in ordinary pulsars 
and MSPs, compared to those in the slowly rotating AXPs/SGRs, may suppress pulsar 
glitches with relatively large enough $\Delta\nu/\nu$ values that can produce 
observable extragalactic FRBs. Note 
that the only two MSP glitches [33] that  were observed so far had 
$\Delta \nu/\nu\!\sim\!10^{-10}$, which implied gravitational energy release of $\Delta 
E_g \!\approx 1.6\times 10^{43}\,{\rm{erg}}$ (see Eq.(7)). Such a gravitational energy 
release at a typical cosmological distance $\gsim\,$Gpc produces an isotropic energy 
fluence below $10^{-10}\, {\rm{erg}/cm^2}$ at Earth. Although such a fluence is below 
the detection thresholds of current x-ray and gamma-ray full sky monitors, such as Swift, 
Konus-Wind, and Fermi GBM, which are above $10^{-8}{\rm{erg}/cm^2}$, it does 
not exclude the possibility that narrowly beamed FRBs without a detectable  
x/$\gamma$-ray flare, which point to Earth, are also detectable from glitches in  
extragalactic MSPs. 

Note however, that  the observational information on glitches in MSPs
is limitrd to old MSPs. Nothing is currently known on glitches in 
very young MSPs, which prevents a defenite conclusion on whether or not 
they, like SGRS/AXPs, are also a main source of extragalactic FRBs.

\section*{Appendix: Magnetars - Myth Or Reality ?}
Magnetars are neutron stars whose dipole magnetic field exceeds $10^{14}$ Gauss 
[28]. Anomalous x-ray pulsars (AXPs) and soft gamma ray repeaters (SGRs), are slowly 
rotating pulsars whose rotational energy loss is too small to power their observed 
x-ray luminosity. They are widely believed to be powered by the decay of their 
estimated huge magnetic field energy. However, their estimated magnetic field is 
based on the {\bf assumption} that they spin down by magnetic dipole radiation. 
Such an assumption yields a polar magnetic field,
\begin{equation}
B^2\sin^2\alpha = {3\,c^3\,I\,P\,\dot{P}\over 2\pi^2\, R^6}\,.
\end{equation}
For a canonical neutron star of a mass $M\approx M_{Ch}$, where
$M_{Ch}$ is the Chandrasekhar mass limit of white dwarfs,
$I\approx (2/5)\, M_{Ch}\, R^2\approx 1.12\times 10^{45}\,{\rm g\, cm^2}$,
Eq.(11) yields a polar magnetic field 
\begin{equation}
B_p\,\sin \alpha \approx 6\times 10^{19}\,[P\dot{P}/s]^{1/2}\, Gauss
\end{equation}
where $\alpha$ is the angle between the magnetic dipole moment and the 
rotation axis. Eq.(12) yields $B_p$ value in excess of $10^{14}$ Gauss 
for most of the known AXPs/SGRs [28].
 
Eq.(12), which has been used widely to establish the magnetar identity of 
SGRs/AXPs [28], is valid only if magnetic dipole radiation (MDR) dominates 
their spin down. However, if the spin down rate of a pulsar by other 
emissions, such as cosmic rays, particle winds, and gravitational waves 
is much larger than that by MDR, then eq.(12) overestimates by far the 
true value of their $B_p$. In fact, the loss of angular momentum of 
AXPs/SGRs can be dominated by emission of highly relativistic charged 
cosmic ray particles. Such cosmic rays gyrate along the magnetic field 
lines and escape when they reach the pulsar's light cylinder of a radius 
$r=c/\omega$ around the pulsar's rotation axis. If the high energy cosmic 
ray (CR) luminosity of a pulsar is $Lum(CR)$, then the loss rate of 
angular momentum by highly relativistic charged cosmic ray particles 
satisfies 
\begin{equation} 
|{\bf \dot{L}}_{CR}|=\dot{n}_i|{\bf 
r\,x\,p}_i|\approx Lum(CR)/\omega, 
\label{Eq13} 
\end{equation} 
where $\dot{n}_i$ is the escape rate of  CR particles from the light 
cylinder and$p_i\!\approx\! E_i/c$ is their momentum. Such a 
cosmic ray luminosity can dominate the spin down of SGRs/AXPs and 
invalidate their estimated magnetic field from  the assumption that MDR 
dominates their spin down. Moreover, the "anomalous" x-ray emission of AXPs 
can be a surface emission powered by gravitational energy release in a 
slow contraction ($R/\dot{R}\gsim 10^{5}$ year).

Moreover, the assumption that a sudden decay (within a few ms) of the 
estimated ultra strong magnetic field of SGRs/AXPs powers 
their giant flares was contradicted by the fact that the product
$P\dot{P}$ SGR 1806-20 has not changed significantly [30] during 
the giant flare of on 27 December 2004 [16]. 

Furthermore, the interpretation of absorption features observed in the 
x-ray spectrum of AXPs/SGRs as due to transitions between Landau 
levels of protons, rather than electrons, in their near surface 
magnetic field [31], results in a magnetic field strength that is 
$(m_p/m_e)\!\approx\! 1830$ times stronger than that obtained for 
electron transitions, which have yielded $B\sim 10^{12}$ Gauss 
typical for ordinary young pulsars. However, the constant magnetic field 
approximation in the pulsar magnetosphere is unreliable. Moreover, the 
absorption cross section for proton transitions between Landau levels 
is suppressed by a factor $(m_e/m_p)^2$ compared to that of electrons, 
as in synchrotron radiation and Compton scattering, which makes the 
proton interpretation very unlikely. 

Note also  that in a couple of cases, the {\bf assumption} that AXPs/SGRs 
spin down by MDR has yielded a surface magnetic field similar to 
that of ordinary pulsars, i.e., much smaller than that expected in 
magnetars. E.g., $B\!<\!\!7.5 \times 10^{12}$ Gauss at the equator
was was obtained for SGR 0418+5729 [32a], and $B\sim 2.7\times 10^{13}$ 
Gauss for SGR 1822-1606 [32b].

Finally, note that the {\bf assumption} that MSPs spin down by MDR has 
yielded a magnetic field strength $B\leq 10^{10}$ Gauss for all MSPs 
with a measured $P/\dot{P}$. However, all these MSPs are rather old 
($ P/2 \dot{P}\geq 10^7$ y), and thus do noy  exclude the possibility 
that very young MSPs can be magnetars, i.e., born with 
a magnetic field which exceeds $10^{14}$ Gauss.


\begin{thebibliography}{}

\bibitem{Lorimer1}
For recent reviews see, e.g., D.R. Lorimer, 2018 [arXiv:1811.00195]; 
J. I. Katz, Prog. Part. Nucl. Phys. 1 {\bf 103}, 1, (2018) 
[arXiv:1804.09092]; E. Petroff, J.W.T. Hessels, D.R. Lorimer,  
A\&A Reviews (2019) [arXiv:1904.07947], and references therein.

\bibitem{Lorimer2}
D. R. Lorimer, M. Bailes, M.A.  McLaughlin, et al.
Science, {\bf 318}, 777 (2007) [arXiv:0709.4301].

\bibitem{Petroff}
E. Petroff, E.D. Barr, A. Jameson, E.F. 
et al. FRBCAT: The Fast Radio Burst Catalogue. PASA 
{\bf 33}, 045 (2016) [arXiv:1601.03547].

\bibitem{Thornton}
D. Thornton,  et al.  Science, {\bf 341}, 53 (2013) [arXiv:1307.1628].
L.G. Spitler, et al., Nature, {\bf 531}, 202 (2016) [arXiv:1603.00581].\\
P. Scholz,  et al.,  ApJ, {\bf 846}, 80 (2017) [arXiv:1705.07824].\\
C.J. Law, et al., ApJ, {\bf 850}, 76 (2017) [arXiv:1705.07553].

\bibitem{Tendulkar}
S.P. Tendulkar, et al., ApJ, {\bf 834}, L7 (2017) 
[arXive:1701.01100].

\bibitem{Fonseca}
E. Fonseca, B.C. Andersen, M. Bhardwaj, et al.
(the CHIME/FRB collaboration) Nature, {\bf 566}, 235 (2019) 
[arXiv:1901.04525].


\bibitem{Andersen}
B.C. Andersen, K. Bandura, M. Bhardwaj, et al., 
(the CHIME/FRB collaboration), ApJ, {\bf 885} L24, 2019 
[arXiv:1908.03507].


\bibitem{Amiri}
M. Amiri, B.C. Andersen, K.M. Bandura, et al., 
(the CHIME/FRB Collaboration) 2020,  e-prints, arXiv:2001.10275.

\bibitem{Platts}
For a catalogue of FRB models see, e.g., 
E. Platts, A. Weltman, A. Walters, et al., [arXiv:1810.05836].

\bibitem{CHIME}
B.C. Andersen, et al., (CHIME/FRB), 
2020, e-print, arXiv:2005.10324.\\
C.D. Bochenek, V. Ravi, K.V. Belov, et al., (STARE2),
Publ. Astron. Soc. Pac. {\bf 132}, 034202 (2020) [arXiv:2005.10828].

\bibitem{INTEGRAL}
S. Mereghetti, V. Savchenko, C. Ferrigno, et al.,
(INTEGRAL),  e-print arXiv:2005.06335 (2020). \\
C.K. Li, L. Lin, S.L. Xiong, M.Y. Ge, et al., 
(Insight-HXMT), e-print arXiv:2005.11071 (2020).\\
A. Ridnai,  D. Svinkin, D. Frederiks, et al.,
(Konus-Wind)  e-print arXiv:2005.11178 (2020)\\
M. Tavani, C. Casentini, A. Ursi, et al., (AGILE)
e-print  arXiv:2005.12164 (2020).

\bibitem{Dado1}
See, for instance, S. Dado \& A. Dar, e-print arXiv:1810.03514,  
and references therein. 

\bibitem{glitches}
See the McGill Magnetar Catalog published by
S.A. Olausen \& V.M. Kaspi, ApJS, {\bf 212}, 6O (2004),
[http://www.physics.mcgill.ca/~pulsar/magnetar/main.html].

\bibitem{AD2}
A. Dar \& A. De R\'ujula,  
Results and Perspectives in Particle Physics (Ed. Mario Greco) Vol. 
XVII (2000) 13 [arXiv:astro-ph/0002014].
 
\bibitem{Colgate}
S.A. Colgate, Canadian Journal of Physics Supplement {\bf 46}, 476 (1968)\\
S.A. Colgate, in Seventh Texas Symposium on Relativistic 
Astrophysics, Annals of the New
York Academy of Sciences, vol. 262, ed. by P.G. Bergman, E.J. 
Fenyves, L. Motz (1975).

\bibitem{Hurley}
K. Hurley, S.E. Boggs, D.M. Smith, et al., 
Nature {\bf 434}, 1098, (2005) [arXiv:astro-ph/0502329].

\bibitem{Kozlova}
A.V. Kozlova, G.L. Israel, D.S. Svinkin, et al.,  MNRAS, {\bf 260}, 408     
(2016) [arXiv:1605.02993].

\bibitem{Curvature}
Bursts of coherent curvature radiation from highly magnetized 
millisecond pulsars have been proposed before as the source of FRBs. 
See, e.g., 
Z.G. Dai, J.S. Wang, X.F. Wu, et al.,  ApJ, {\bf 829}, 27 (2016)
[arXiv:1803.09945];
J.S. Wang, Y.P. Yang, X.F. Wu, et al. ApJ, {\bf 822}, L7 (2016)
[arXiv:1603.02014]; 
W.M. Gu, Y.Z. Dong, T. Liu, et al., ApJ, {\bf 670}, 693 (2016) 
[arXiv:1604.05336];
G. Ghisellini, N. Locatelli, A\&A,  {\bf 613}, 61 (2017)
[arXiv:1708.07507];
J. Katz, MNRAS, {\bf 481}, 2946 (2018) 
[arXiv:1803.01938];
Y.P. Yang \& B. Zhang 2017 
[arXiv:1712.02702];
P. Kumar, W. Lu, M. Bhattacharya, MNRAS, {\bf 468}, 2726 (2017)
[arXiv:1703.06139]\\
The discovery that the Galactic SGR 1935$+$2154 emitted FRB 200428
led J. Katz [arXiv:2006.03468] to suggest that coherent curvature 
radiation emitted in giant flares of SGRs/AXPs, which are widely 
believed to be sllowly rotating magnetars, produce FRBs.
While the observational data seems to support the idea that  
FRBs are coherent curvature radiation emitted in large flares 
of SGRs/AXPs, there is no compelling 
observational evidence that their steady x-ray emission and/or 
flaring activity are powered by the decay of their 
{\bf assumed} ultra strong magnetic field (see the Appendix
of this paper).

\bibitem{Michilli}
D. D. Michilli, A. Seymour,  J.W.T. Hessels,  et al.,
Nature, {\bf 553}, 182 (2018) [arXiv:1801.03965].\\
M. Caleb, E.F. Keane, W. van Straten, et al. 
e-print arXive:1804.09178 (2018) and references therein.\\ 
C.K. Day, A.T. Deller, R.M. Shannon, e-print arXiv:2005.13162 (2020).

\bibitem{Manchester1}
V. Radhakrishnan \&R.N. Manchester, Nature, {\bf 222}, 228 (1969).\\ 
P.E. Reichley \& G.S. Downs, Nature, {\bf 222}, 229 (1969).\\
For a recent review see, R.N. Manchester in
Pulsar Astrophysics - The Next 50 Years, Cambridge University Press,
[arXiv:1801.04332] and references therein.\\
For summary of glitches in pulsars, see, e.g.,\\ 
http://www.jb.man.ac.uk/pulsar/glitches.html

\bibitem{Migdal}
A.B. Migdal, Soviet Physics JETP, {\bf 10}, 176 (1960).

\bibitem{Pines}
D. Pines, J. Shaham, M. A. Ruderman, Nature, {\bf 237},
83 (1972).\\
P. W. Anderson and N. Itoh, N. Nature, {\bf 256}, 25 (1975).\\
This prevailing theory of pulsar glitches 
has been called into question several times. See, e.g.,\\
N. Andersson, K. Glampedakis, W. Ho, C. Espinoza, 
PRL, {\bf 109}, 241103 (2012) [arXiv:1207.0633].
See, however, B. Haskell, A. Melatos, IJMP, 
IJMPD 24, issue 3, (2015) [arXiv:1502.07062]. 

\bibitem{MSP}
Only two glitches in millisecond pulsars, 
PSRs B1821-24A and J0613-0200, have been observed, so far.
Each  had just one very small glitch with
$\Delta\nu/\nu\!\approx\! 10^{-11}$.\\
I. Cognard, D. C. Backer,  ApJ, {\bf 612}, L125 (2004) 
[arXiv:astro-ph/0407546].\\
J.W. McKee, G.H.,  Janssen,  B.W.  Stappers,  et al. MNRAS
{\bf 461}, 2809 (2016) [arXiv:1606.04098]. 

\bibitem{Bibby}
J.L. Bibby, P.A. Crowther, J.P. Furness, J.S. Clark,
MNRAS {\bf 386}, L23, 2008 [arXiv:0802.0815]. 

\bibitem{Zhou}
P. Zhou, X. Zhou, Y. Chen, J. Wang, et al. 
e-print arXiv:2005.03517 (2020).

\bibitem{Jackson}
J.D. Jackson, Classical Electrodynamics (3rd ed.). Chichester: Wiley 
(1999).

\bibitem{Risk}
K. Hurley, S.E. Boggs, D.M. Smith, et al.,
Nature {\bf 434}, 1098, (2005) [arXiv:astro-ph/0502329].

\bibitem{Magnetrs}
Originally, magnetars were defined to be 
pulsars with an ultra strong magnetic field whose decay powers their 
radiation [R.C. Duncan \& C. Thompson, ApJ {\bf 392}, L9 (1992): R.C. 
Duncan, ApJ, {\bf 408}, 194 (1993)]. The slowly rotating anomalous 
x-ray pulsars (AXPs) and soft gamma ray repeaters (SGRs), whose 
observed x-ray luminosity was found to exceed their loss rate of 
rotational energy [see, e.g., S. Mereghetti, A\&AR, {\bf 15}, 225 
(2008) [arXiv:0804.0250]] were the first type of pulsars which were 
claimed to be magnetars [C. Kouveliotou., S. Dieters., T. Strohmayer., 
et al. Nature, {\bf 393}, 235 (1998)]. Their spin down was assumed to be 
powered by magnetic dipole radiation (MDR) while their steady x-ray emission
and x/$\gamma$-ray flares by the decay of their huge magnetic 
field energy (see, e.g., V.M. Kaspi, \& A.M. Beloborodov, Annu. Rev. 
Astron. Astrophys. {\bf 55}, 261 (2017) [arXiv:1703.00068] for a 
recent review, and references therein).

\bibitem{Manchester}
R. N. Manchester \& J.H. Taylor,  Pulsars (W. H.
Freeman \& Company, San Francisco, 1977).

\bibitem{Woods}
P.M. Woods, C. Kouveliotou, M.H. Finger, et al., ApJ, {\bf 654}, 470 (2006) 
[arXiv:astro-ph/0602402].

\bibitem{Ibrahim}
A.I. Ibrahim, S. Safi-Harb, J.H. Swank, et al., ApJ, {\bf 574}, L51 (2002) 
[arXiv:astro-ph/0210513].

\bibitem{Rea}
(a) N. Rea, P. Esposito, R. Turolla, et al., Science, {\bf 330}, 994 
(2010) [arXiv:1010.2781].\\
(b) N. Rea, G.L. Israel, P. Esposito, et al., ApJ, {\bf 754}, 27 
(2012)[arXiv:1211.7347].

\bibitem{Cognard}
I. Cognard \& D.C. Backer, ApJ, 612, L125 (2004) [arXiv:astro-ph/0407546].\\
J.W. McKee, G.H., Janssen, B.W. Stappers et al.  MNRAS {\bf 461}, 2809 
(2016) [arXiv:1606.04098].

\end{thebibliography}
\end{document}